\documentclass[stef,twoside]{stefano}

\usepackage{amsmath}
\usepackage{amssymb}
\usepackage{graphics}
\usepackage{rotating}
\usepackage{cite}
\usepackage{color}

\def\makeheadbox{{%
\hbox to0pt{\vbox{\baselineskip=10dd\hrule\hbox
to\hsize{\vrule\kern3pt\vbox{\kern3pt \hbox{{\sc European Physics
Journal C} {\bf 53}, 659-666 (2008) }
\hbox{
{\sc {\color{blue}{dma}}[{\color{black}{imecc}}]{\color{red}{UniCamp}}
}
\hspace*{10.3cm}
{\color{blue}{$\boldsymbol{\Sigma \delta \Lambda}$ }}}
\kern3pt}\hfil\kern3pt\vrule}\hrule}%
\hss}}}

\def\0{\mbox{\tiny $0$}}
\def\1{\mbox{\tiny $1$}}
\def\2{\mbox{\tiny $2$}}
\def\3{\mbox{\tiny $3$}}
\def\4{\mbox{\tiny $4$}}
\def\5{\mbox{\tiny $5$}}
\def\6{\mbox{\tiny $6$}}
\def\7{\mbox{\tiny $7$}}
\def\8{\mbox{\tiny $8$}}
\def\9{\mbox{\tiny $9$}}

\def\eff{\mbox{\tiny eff}}
\def\tree{\mbox{\tiny tree}}
\def\Y{\mbox{\tiny Y}}
\def\H{\mbox{\tiny H}}
\begin{document}
%

\title{\Large BOUND STATE INEQUALITY FOR HIGH MASS EXCHANGES IN A
SCALAR FIELD MODEL}

\author{
Stefano De Leo \inst{1}
\and Pietro Rotelli\inst{2} }

\institute{
Department of Applied Mathematics, University of Campinas\\
PO Box 6065, SP 13083-970, Campinas, Brazil\\
{\em deleo@ime.unicamp.br}\\
 \and
Department of Physics, University of
Lecce and INFN Lecce\\
PO BOX 193, CAP 73100, Lecce, Italy\\
{\em rotelli@le.infn.it} }


\date{Submitted {\em July, 2007}}

\abstract{Ladder diagrams are relevant for the study of bound states. The
condition upon the coupling strength for the existence of a bound state has
been deduced in a scalar field theory for the case of low mass exchanges.
We apply this approach to the case of very high mass exchanges.}















\titlerunning{\sc bound states in quantum field theory}

\maketitle


\section*{I. INTRODUCTION}

A fundamental question in quantum mechanics is the existence, spectrum and
properties of bound states\cite{COHEN,ZUB,SAK,GROSS}. The importance of
this subject matter for atomic and molecular physics is obvious. In
particle physics it is also of primary importance. Positronium, heavy quark
bound states such as $J/\psi$ and indeed all quark (anti-quark) bound
states are of interest to physicists and often provide information about
constituent roles and theoretical models. Resonances are a related subject
matter but unlike bound states they are above threshold and have finite
lifetimes either because they decay via a weaker interaction, such as for
the flavor changing quark decays, or because they are ``excited'' states
that eventually de-excite to the ground state\cite{ZUB}.

The main approach to bound states has been the solution of the appropriate
non-relativistic wave equation in the presence of a
potential\cite{ZUB,SAK,GROSS}. This is perfectly adequate when the
solutions yield non-relativistic bound states. For better precision, their
relativistic corrections can be treated with equations such as the Dirac or
Klein-Gordon equations. However, these relativistic equations exhibit one
very important limitation\cite{AYC} when used for calculating bound states,
related to the Klein paradox\cite{KLE29,HR81,SMG82,KSG04,DRKP}.

For long range interactions, such as electromagnetic or gravitational, an
infinite number of bound states exist. For short range interactions such as
that given by a Yukawa potential,
\begin{equation}
V_{\Y}(r)=-\,\frac{g_{\eff}^{\2}}{4\,\pi}\, \frac{\exp[-\,\mu \,r]}{r} \,\,,
\end{equation}
corresponding to the exchange of a particle of mass $\mu$, only a finite
number of bound states exist if any at all. Indeed for small mass exchanges
(compared to the systems reduced mass), numerical calculations within the
Schr\"odinger equation yield a minimum condition for the existence of a
bound state, i.e.
\begin{equation}
\label{condg} \frac{g^{\2}_{\eff}}{4\,\pi} \ge 0.84\, \frac{\mu}{m}\,\,,
\end{equation}
where $\mu$ is the exchanged particle mass and $m$ is the reduced mass. To
the extent that this inequality is valid, i.e. that the potential is
realistic and that the Schr\"odinger equation is acceptable, it tells us
that as the range of the interaction falls ($\mu$ and hence $\mu/m$
increases) the coupling strength must grow as $\mu/m$ to permit the
existence of a bound state. However, as the coupling grows so does, in
general, the binding energy, $E_b$. For example, in the Bohr model $ E_b
\propto \alpha^{\2}$, with $\alpha\equiv g^{\2}_{\eff}/4\,\pi$,
 so that eventually the Schr\"odinger
equation becomes inappropriate  as the binding energy tends to or exceeds
the reduced mass.

If one considers a potential which is akin to the Yukawa potential
but has the advantage of being solvable exactly\cite{FLU}, i.e.
the Hulthen potential \begin{equation}
V_{\H}=-\,\frac{g_{\eff}^{\2}}{4\,\pi}\,
\frac{2\,\mu}{\exp[2\,\mu\, r] -1}\,\,, \end{equation}
 one finds, {\em analytically},  the condition for the single (s-wave) bound state to exist to be
 similar to the numerically inequality found in Eq.(\ref{condg}), i.e.,
\begin{equation}
 \frac{g^{\2}_{\eff}}{4\,\pi} \ge
\frac{\mu}{m}\,\,.\end{equation}
 For completeness, we should point out
 that there is a whole class of Hulthen potentials with $\mu$ in the above
 expression substituted by $c\,\mu$  (where $c$ is a positive constant).
 Our particular choice ($c=1$) is that which shares with the  Yukawa (mass $\mu$)
 the same first two terms in a Maclaurin series expansion about $r=0$.

 Relativistic spinor bound states, by which we mean any spinor bound states
 for which relativity plays an important role, require more care.
 We can first attempt to treat these
states by including the lowest order relativistic corrections to
the Schr\"odinger equation or by passing directly to the Dirac
equation\cite{COHEN,ZUB,SAK,GROSS} . In either of these cases, one
can demonstrate that as $\mu$ increases relativistic effects
automatically increase the effective coupling constant. We have
called this effect the {\em amplification of the Yukawa
coupling}\cite{AYC}. This opens the practical possibility of high
mass exchange bound states. However, if the bound state energy
grows with the effective coupling constant, as numerical
calculations suggest, then we will eventual enter the so-called
Klein zone $E<V_{\0}-m$. Conventionally $V_{\0}$, which is the
asymptotic free space value, is set to zero. Within the Klein zone
only oscillatory solutions exist everywhere. This is the origin of
the Klein paradox which can be interpreted as a consequence of
pair creation\cite{KLE29,HR81,SMG82,KSG04,DRKP}. This is a
positive feature, if considered an anticipation of field theory,
but it is a problem for the one-particle interpretation of these
equations. Furthermore, the absence of evanescent solutions means
the absence of any (discrete spectrum) bound states.

At this point, one naturally passes to field theory. This seems promising
since one of the greatest successes of renormalized field theory is the
calculation of the Lamb shift\cite{LAMB}. Unfortunately the {\em very
existence} of a bound state, while a more elementary question, seems much
more difficult to answer in field theory. This is the reason that one often
falls back upon (heuristic) two-body relativistic equations, albeit
inspired by field theory, such as the Bethe-Salpeter
equation\cite{BS1,BS2,BS3}, the Blankenbecler-Sugar equation\cite{BSug} or
the Gross (spectator) equation\cite{GROSS}. There is however one technique,
described in detail by Franz Gross\cite{G69}, which offers us a very useful
tool. This is based upon the sole consideration of ladder diagrams and
works impressively for small $\mu/m$. Indeed, after introduction of the
scalar model in the next section, we present in section III a simplified
(zero momentum) calculation (for light mass exchanges) which exactly
reproduces the Hulthen inequality. In section IV, we consider the opposite
limit where $\mu/m$ is very large. We seek the appropriate inequality
condition for the existence of a bound state for this limit and, with the
help of some numerical calculations, this will indeed be found.

The  reasons for our interest in this {\em high mass exchange
limit} is that some physical interactions do indeed involve very
heavy mass exchanges\cite{ZUB}, e.g. the exchange of  of the
intermediate vector bosons $W^{\pm}$ and $Z^{\0}$, and almost
certainly of the Higgs particle. In particular, since the neutrino
is now known to have mass eigenstates\cite{PDG}, it is a
legitimate question to ask if the weak interactions allow for,
say, neutrino-lepton bound states\cite{AYC}. The above
inequalities suggests not, for we are asking if a bound state can
exist with a $\mu/m$ ratio $>10^{10}$ for neutrino-electrons
(although this ratio could be much smaller for the heavier
leptonic families). However, these inequalities have been derived
from non-relativistic equations or, as we shall see, from ladder
diagrams in which the assumption of small $\mu/m$ is made from the
start. We shall return briefly to this discussion in our
conclusions.

\section*{II. THE SCALAR MODEL}

Let us consider a scalar model with three different mass scalars. Two of
them represent the incoming system and have mass $m_{\1}$ and  $m_{\2}$.
They interact only by the exchange of a third scalar with mass $\mu$. The
dimensional coupling constants are $\lambda_{\1}$ and $\lambda_{\2}$ for
the particle with mass $m_{\1}$ and  $m_{\2}$ respectively. The one boson
exchange diagram gives the lowest order contribution to the invariant
scattering amplitude,
\begin{equation}
\label{tree} \mathcal{M}_{\tree}(q) = i\,(-\,i\,\lambda_{\1})\,(-\,i\,
\lambda_{\2})\,\frac{-\,i}{\mu^{\2}-q^{\2}-i\,\epsilon} =-\,
\frac{\lambda_{\1}\lambda_{\2}}{\mu^{\2}-q^{\2}-i\,\epsilon}\,\,.
\end{equation}
From the Fourier transform of Eq.(\ref{tree}) when $q_{\0}=0$  one obtains
the Yukawa potential quoted above. The corresponding force is attractive
(and hence can yield bound states) only if $\lambda_{\1} \lambda_{\2} > 0$.
In fact, comparing with the attractive Yukawa shows that
\begin{equation}
\label{geff}
g_{\eff}^{\2}=\frac{\lambda_{\1}\lambda_{\2}}{4\,m_{\1}\,m_{\2}}\,\,.
\end{equation}
Henceforth this is what we shall assume throughout. The scattering
amplitude reduces to a number for forward scattering,
\begin{equation}
\label{tree0} \mathcal{M}_{\tree}(0) = - \,
\frac{\lambda_{\1}\lambda_{\2}}{\mu^{\2}}\,\,.
\end{equation}
The fourth order ladder and crossed (ladder) diagrams are shown in
Fig.\,1. Not surprisingly, these contributions modify the Yukawa
potential as do all higher order terms. We shall come back to this
discussion in the next section. The box and crossed box diagrams
are not the only fourth order diagrams, but the others can be
absorbed into the dressing of the propagators and vertex
functions. As a consequence of the latter, we expect the
appearance of form factors which however reduce to unity for
forward scattering. For our purposes it is sufficient to limit our
calculations to forward scattering. Consequently, we will not
consider explicitly these other diagrams \cite{GROSS}.

In the center of mass system and for forward scattering (see Fig.\,1), the
Feynman rules for the box diagram amplitude yield
\begin{equation}
\mathcal{M}^{\square}=i\,\lambda_{\1}^{\2}\lambda_{\2}^{\2}
\int\frac{\mbox{d}^{\4}k}{(2\pi)^{\4}}\,\frac{1}{D_{\1}D_{\2}D^{^{\2}}_{\0}}
\end{equation}
with
\begin{eqnarray}
D_{\1} & = & \left[\,E_{\1}(\boldsymbol{k})-
E_{\1}(\boldsymbol{p})-E_{\2}(\boldsymbol{p})+k_{\0}-i\epsilon \right]\,
\left[\,E_{\1}(\boldsymbol{k})+
E_{\1}(\boldsymbol{p})+E_{\2}(\boldsymbol{p})-k_{\0}-i\epsilon \right]\,\,,
\nonumber \\
D_{\2} & = & \left[\,E_{\2}(\boldsymbol{k})+k_{\0}-i\epsilon\right]\,
\left[\,E_{\2}(\boldsymbol{k})-k_{\0}-i\epsilon\right]\,\,, \nonumber \\
D_{\0} & = & \left[\,E_{\0}(\boldsymbol{k}-\boldsymbol{p})-
E_{\2}(\boldsymbol{p})+k_{\0}-i\epsilon \right]\,
\left[\,E_{\0}(\boldsymbol{k}-\boldsymbol{p})+
E_{\2}(\boldsymbol{p})-k_{\0}-i\epsilon \right]\,\,,
\end{eqnarray}
and \[
E_{\1,\2}(\boldsymbol{q})=\sqrt{\boldsymbol{q}^{\2}+m_{\1,\2}^{\2}}\,\,\,,
\,\,\,\,\,E_{\0}(\boldsymbol{q})=\sqrt{\boldsymbol{q}^{\2}+\mu^{\2}}
\,\,.\]
Evaluating the propagators near threshold ($\boldsymbol{p}\approx
\boldsymbol{0}$), we find
\begin{eqnarray}
D_{\1} & = & \left[\,E_{\1}(\boldsymbol{k})- m_{\1}-m_{\2}+k_{\0}-i\epsilon
\right]\, \left[\,E_{\1}(\boldsymbol{k})+ m_{\1}+m_{\2}-k_{\0}-i\epsilon
\right]\,\,,
\nonumber \\
D_{\2} & = & \left[\,E_{\2}(\boldsymbol{k})+k_{\0}-i\epsilon\right]\,
\left[\,E_{\2}(\boldsymbol{k})-k_{\0}-i\epsilon\right]\,\,, \nonumber \\
D_{\0} & = & \left[\,E_{\0}(\boldsymbol{k})- m_{\2}+k_{\0}-i\epsilon
\right]\, \left[\,E_{\0}(\boldsymbol{k})+ m_{\2}-k_{\0}-i\epsilon
\right]\,\,.
\end{eqnarray}
For the crossed box diagram (see Fig.\,1) only the internal propagator for
particle with mass $m_{\1}$ has a different momentum. Consequently,
\begin{equation}
\mathcal{M}^{\times}=i\,\lambda_{\1}^{\2}\lambda_{\2}^{\2}
\int\frac{\mbox{d}^{\4}k}{(2\pi)^{\4}}\,\frac{1}{D^{\times}_{\1}D_{\2}
D^{^{\2}}_{\0}}
\end{equation}
with
\begin{eqnarray}
D^{\times}_{\1} & = & \left[\,E_{\1}(\boldsymbol{k})- m_{\1}+
m_{\2}+k_{\0}-i\epsilon \right]\, \left[\,E_{\1}(\boldsymbol{k})+
m_{\1} - m_{\2}-k_{\0}-i\epsilon \right]\,\,.
\end{eqnarray}
The box and crossed  diagrams contain eight poles each in the
complex $k_{\0}$ plane. Half lie below the real axis and
contribute to the integral if we close the contour in the lower
half plane. For the box and crossed diagrams the residues will be
labelled $R^\square_{\0,\1,\2}(\boldsymbol{k})$ and
$R^\times_{\0,\1,\2}(\boldsymbol{k})$ respectively. There are only
three residues, and not four, because for forward scattering in
the rest frame limit the two poles in the exchanged particle
propagators coincide and yield the "double pole" residues
$R^{\square,\times}_{\0}$. Thus, the box and crossed diagrams give
the following fourth order contribution to the invariant
scattering amplitude
\begin{eqnarray}
\label{integral} \mathcal{M}^{\square}+\mathcal{M}^{\times} &=&
i\,(-2\,\pi\,i)\,
\frac{\lambda_{\1}^{^{\2}}\lambda_{\2}^{^{\2}}}{(2\pi)^{^{\4}}}\,
\int\,\mbox{d}^{\3}k \, \sum_{s=\0}^{\2} \left[\,
R_{s}^{\square}(\boldsymbol{k})+R_{s}^{\times}(\boldsymbol{k})\,\right] \nonumber \\
 &  = &
 \frac{\lambda_{\1}^{^{\2}}\lambda_{\2}^{^{\2}}}{2\,\pi^{^{\2}}}\,
\int_{\0}^{\infty}\mbox{d}k \,\,k^{\2}\,\,\sum_{s=\0}^{\2}
\left[\, R_{s}^{\square}(k)+R_{s}^{\times}(k)\,\right]\,\,.
\end{eqnarray}
 Below by $E_s$ we intend $E_s(k)$ and by $W$
and $\Delta$ we intend $m_{\1}+m_{\2}$ and $m_{\2}-m_{\1}$
respectively. A simple calculation shows that the explicit
formulas for the residues in the $k_{\0}$-plane for the box and
the crossed diagrams are respectively
\begin{eqnarray}
R_{\1}^{\square}(k)
 & = & 1\,/\,\left\{\,4\, W\,E_{\1}
 \left(E_{\1}+m_{\1}\right)\left[\,\mu^{\2}-2\,m_{\1}\left(E_{\1}+m_{\1}\right)\,
\right]^{\2}\right\}\,\,, \nonumber \\
R_{\2}^{\square}(k)
 & = & -\,1\,/\,\left\{\,4\, W\,E_{\2}
 \left(E_{\2}-m_{\2}\right)\left[\,\mu^{\2}+2\,m_{\2}\left(E_{\2}-m_{\2}\right)\,
\right]^{\2}\right\}\,\,,\\
R_{\0}^{\square}(k) & = & \left[\,2\,\left(E_{\0}-m_{\1}\right)\,B\,C +
2\,\left(E_{\0}+m_{\2}\right)\,A\,C -2\,A\,B\,\right]\,/\,
\left[\,A^{\2}B^{\2}C^{\3}\,\right]\,\,, \nonumber
\end{eqnarray}
with
\[A=2\,E_{\0}\,m_{\1} - \mu^{\2}\,\,\,,\,\,
B=- 2\,E_{\0}\,m_{\2} - \mu^{\2}\,\,\,\,\,\mbox{and}\,\,\,\,\,
C=2\,E_{\0}\,\,,
\]
and
\begin{eqnarray}
R_{\1}^{\times} & = & 1\,/\,\left\{\,4\, \Delta\,E_{\1}
 \left(E_{\1}-m_{\1}\right)\left[\,\mu^{\2}+2\,m_{\1}\left(E_{\1}-m_{\1}\right)\,
\right]^{\2} \right\}\,\,,\nonumber \\
R_{\2}^{\times}
 & = & -\,1\,/\,\left\{\,4\, \Delta\,E_{\2}
 \,\left[\,\mu^{\2}+
2\,m_{\2}\left(E_{\2}-m_{\2}\right)\,
\right]^{\2}\right\}\,\,,\\
R_{\0}^{\times} & = & \left[\,2\,\left(E_{\0}+m_{\1}\right)\,B\,C +
2\,\left(E_{\0}+m_{\2}\right)\,A_{\times}\,C -
2\,A_{\times}\,B\,\right]\,/\,\left[\,A_{\times}^{\2}B^{\2}C^{\3}\,\right]\,\,,
\nonumber
\end{eqnarray}
with
\[A_{\times}=-\,\left(2\,E_{\0}\,m_{\1} + \mu^{\2}\right)\,\,.
\]
We warn the reader that our choice of labelling of momenta for the
crossed diagram is different from that of Franz
Gross\cite{GROSS,G69}. This results in different contributions
from the various crossed poles. Of course, the sum over the poles
gives the same result (see the next section).

Franz Gross conjectures in his classical book on relativistic
quantum mechanics\cite{GROSS} that the {\em inequality condition}
for a bound state to exist can be derived by equating the
contributions of the tree and box diagrams. Actually, for a bound
state one expects the perturbation series to diverge and, in
particular, for each order in the ladder series to be of
comparable strength. In this paper, we will limit ourselves to the
much simpler task of comparing the second order three amplitude to
the fourth order box terms.

\section*{III. THE EXCHANGE OF SMALL MASS SCALARS}

For incoming scalars with mass $m_{\1}$ and $m_{\2}$ interacting by the
exchange of a third scalar with mass $\mu \ll m_{\1,\2}$, the integrand
functions which appear in Eq.(\ref{integral}), i.e.
\[ k^{\2}R^{\square}(k) =k^{\2} \sum_{s=\0}^{\2} R_s^{\square}(k)\,\,\,\,\,
\mbox{and}\,\,\,\,\, k^{\2}R^{\times}(k) =k^{\2} \sum_{s=\0}^{\2}
R_s^{\times}(k)\,\,, \] contribute to the invariant scattering amplitude
only for value of $k\ll m_{\1,\2}$. The $k$-dependence of these functions
is explicitly shown, for particular values of $m_{\1,\2}$ and $\mu$ in the
case (b) of Fig.\,2. In this small $\mu$ limit,
\[
\frac{R_{\1}^{\square}(k)}{R_{\2}^{\square}(k)} =  -\,
\frac{E_{\2}-m_{\2}}{E_{\1}+m_{\1}}\, \frac{E_{\2}}{E_{\1}}\,
\left[\,\frac{\mu^{\2}+2\,m_{\2}\left(E_{\2}-m_{\2}\right)}{
\mu^{\2}-2\,m_{\1}\left(E_{\1}+m_{\1}\right)}\,\right]^{\2} \approx
-\,\frac{k^{\2}}{4\,m_{\1}^{\2}}\, \left(\, \frac{k^{\2}+ \mu^{\2}}{k^{\2}+
4\,m_{\1}^{\2}- \mu^{\2}}\,\right)^{\2}\,\,.
\]
This is a very small ratio, so $R_{\2}^{\square}(k)$ dominates. This
residue can be approximated by
\[
k^{\2}\,  R_{\2}^{\square}(k)  =  -\,(E_{\2}+m_{\2})\,/\,\left\{\,4\,
W\,E_{\2}\,\left[\,\mu^{\2}+2\,m_{\2}\left(E_{\2}-m_{\2}\right)
\right]^{\2}\right\}  \approx   -\,1\,/\,\left[\,2\, W\,
 \left(k^{\2}+\mu^{\2}
\right)^{\2}\,\right]\,\,,
\]
and, by making use of the elementary integrals
\[ \alpha^{\3}\,\int_{\0}^{\infty} \frac{\mbox{d}k}{(k^{\2}+\alpha^{\2})^{^{\2}}}
=\alpha\,\int_{\0}^{\infty}
\frac{k^{\2}\,\mbox{d}k}{(k^{\2}+\alpha^{\2})^{^{\2}}}=
\frac{\pi}{4}\,\,,\] we find that
\begin{equation}
\int_{\0}^{\infty}\mbox{d}k\,\,k^{\2}\,R_{\2}^{\square}(k) \approx
-\,\frac{\pi}{8\,W\,\mu^{\3}}\,\,,
\end{equation}
which, as anticipated, is much larger than the contribution of
$R_{\1}^{\square}(k)$ given by
\begin{equation}
\int_{\0}^{\infty}\mbox{d}k\,\,k^{\2}\,R_{\1}^{\square}(k) \approx
\frac{\pi}{64\,W\,m_{\1}^{\3}}\,\,.\end{equation}
 The corresponding crossed
contributions are
\begin{eqnarray*}
k^{\2}\,R_{\1}^{\times}(k) & = &
\left(E_{\1}+m_{\1}\right)\,/\,\left\{\,4\, \Delta\,E_{\1}
\left[\,\mu^{\2}+2\,m_{\1}\left(E_{\1}-m_{\1}\right)
\right]^{\2} \right\}\\
&\approx&  1\,/\,\left[\,2\, \Delta\,
 \left(k^{\2}+\mu^{\2}
\right)^{\2}\,\right] + k^{\2}\,/\, \left[\,8\,m_{\1}^{\2}\, \Delta\,
 \left(k^{\2}+\mu^{\2}
\right)^{\2}\,\right]  \,\,,\\
k^{\2}\,R_{\2}^{\times}(k) & = & -\,
\left(E_{\2}+m_{\2}\right)\,/\,\left\{\,4\, \Delta\,E_{\2}
 \, \left[\,\mu^{\2}+ 2\,m_{\2}\left(E_{\2}-m_{\2}\right)
\right]^{\2}\right\}\\
& \approx & -\,1\,/\,\left[\,2\, \Delta\,
 \left(\,k^{\2}+\mu^{\2}
\right)^{\2}\,\right]- k^{\2}\,/\, \left[\,8\,m_{\2}^{\2}\, \Delta\,
 \left(k^{\2}+\mu^{\2}
\right)^{\2}\,\right]\,\,.
\end{eqnarray*}
In these expressions for the crossed residues we have kept higher order
terms because the leading contributions cancel. In fact,
\[ k^{\2}\,\left[\,R_{\1}^{\times}(k) +R_{\2}^{\times}(k)\,\right]
\approx \frac{W}{8\,m_{\1}^{\2}\,m_{\2}^{\2}}\,
\frac{k^{\2}}{\left(k^{\2}+\mu^{\2}\right)^{\2}} \,\,.\] and hence
\begin{equation}
\int_{\0}^{\infty}\mbox{d}k\,\,k^{\2}\,\left[\,R_{\1}^{\times}(k)
+R_{\2}^{\times}(k)\,\right] \approx \frac{\pi}{32\,W\,m^{\2}\,\mu}\,\,,
\end{equation}
where $m=m_{\1}m_{\2}/W$ is, as before, the reduced mass.

Let us now consider the "double pole" contributions. Since both $k^{\2}$
and $\mu^{\2}$ are very small compared to $m_{\1,\2}^{\2}$, we can
approximate the expressions for $A$, $A_{\times}$ and $B$ by
\[ A/m_{\1}\approx -\,A_{\times}/m_{\1} \approx -\,B/m_{\2}
\approx C =2\,E_{\0}\,\, , \]
whence
\[ R_{\0}^{\square}(k) \approx -\, R_{\0}^{\times}(k)  \approx
3\,k^{\2}\,/\,\left[\,16\,m_{\1}m_{\2}\, \left(\,\mu^{\2}+k^{\2}
\right)^{^{\5/\2}}\,\right]\,\,.
\]
Now with the help of the elementary integral
\[  \alpha^{\2}\,\int_{\0}^{\infty}
\frac{k^{\2}\,\mbox{d}k}{(k^{\2}+\alpha^{\2})^{^{\5/\2}}} =
\frac{1}{3}\,\,,\] we find that
\begin{equation}
\int_{\0}^{\infty}\mbox{d}k\,\,k^{\2}\,R_{\0}^{\square} (k)\approx -\,
\int_{\0}^{\infty}\mbox{d}k\,\,k^{\2}\,R_{\times}^{\square}(k)
 \approx
\frac{1}{16\,W\,m\,\mu^{\2}}\,\,.\end{equation} Finally, the contributions
to the scattering amplitude, coming from the fourth order ladder and
crossed ladder diagrams, can be analytically expressed by using the leading
contributions coming from the single pole $2$ and the double pole $0$ for
the box diagram, i.e.
\begin{equation}
\label{ts1} \mathcal{M}^{\square} \approx
\frac{\lambda_{\1}^{^{\2}}\lambda_{\2}^{^{\2}}}{2\,\pi^{\2}}\, \left[\,
-\,\frac{\pi}{8\,W\,\mu^{\3}} + \,\frac{1}{16\,W\,m\,\mu^{\2}} \right]\,\,,
\end{equation}
 and from the double pole $0$ for the crossed one, i.e.
\begin{equation}
\label{ts2} \mathcal{M}^{\times} \approx
\frac{\lambda_{\1}^{^{\2}}\lambda_{\2}^{^{\2}}}{2\,\pi^{\2}}\, \left[\, -
\,\frac{1}{16\,W\,m\,\mu^{\2}} \right]\,\,.
\end{equation}
These analytic expressions are in excellent agreement with
numerical (test) calculations, made for a selected choice of
masses. These numerical results have been obtained by using
directly Eq.(\ref{integral}), i.e. without any approximations.
This is shown in the upper part of Table 1, that for $\mu\ll
m_{\1,\2}$, in which by "analytic box" and "analytic crossed" we
intend the expressions in the brackets of Eqs.(\ref{ts1}) and
(\ref{ts2}).

Comparing now the fourth-order total scattering amplitude,
 \begin{equation}
\mathcal{M}^{\square}+\mathcal{M}^{\times} \approx - \,
\frac{\lambda_{\1}^{^{\2}}\lambda_{\2}^{^{\2}}}{16\,\pi}\,
\frac{1}{W\,\mu^{\3}}\,\,,
\end{equation}
with the one boson exchange amplitude (\ref{tree0}), we find that the
fourth-order amplitude is greater or comparable to the second-order
amplitude when
\begin{equation}
\label{cond} \frac{\lambda_{\1}\lambda_{\2}}{16\,\pi\,W\,\mu} \geq
1\,\,.\end{equation} By using the {\em effective dimensionless} coupling
strength for the $\Phi^{\3}$ Yukawa interaction, see Eq.(\ref{geff}), the
previous condition becomes
\begin{equation}
\frac{g_{\eff}^{^{\2}}}{4\pi} \geq \frac{\mu}{m}  \,\,,
\end{equation}
which reproduces exactly the Hulthen inequality given in section I.

 The fourth order terms considered modify significantly the "effective"
potential in the calculation. As shown by Gross\cite{GROSS} the added
potential to the tree diagram Yukawa is given by
\[ V_{\2 \mu}(r) =
\frac{1}{8\,\pi}\,\left(\,\frac{g^{\2}_{\eff}}{4\,\pi}\,\right)^{\2}\,\int_{\2
\mu}^{\infty}\,\frac{\mbox{d}z}{\sqrt{z^{\2}-4\,\mu^{\2}}}\,\frac{\exp[-\,z\,r]}{r}
\,\,.\] This represents an integral over higher mass ($>2\mu$)
exchanges. It implies a significant addition to the Yukawa. Higher
order terms will also produce modifications. We expect that the
basic (underlying) Yukawa should  become insignificant  as we
approach the bound state inequality, after which the perturbation
series diverges. If the Yukawa is indeed "smothered" out, it is
somewhat surprising  that the above bound state inequality is
exactly the same as that given by the non-relativistic Hulthen.

Finally, there is an important point, made by Gross\cite{G69}, that we wish
to recall about this approach. The perturbation series (ladder diagrams)
considered are relativistically invariant. For small $\mu$ the loop
momentum is also small, on average, and consequently the relativistic
corrections are small. These corrections are associated principally with
the double pole contributions. However, for this {\em particular model},
the double pole contributions of the box and crossed diagrams cancel to
leading order (see the above approximate equations). This observation will
be relevant for our conclusions.

\section*{IV. THE EXCHANGE OF HIGH MASS SCALARS}

We now proceed to the original part of this work. We consider the case of
large $\mu$ exchange, i.e. when $\mu \gg m_{\1,\2}$. We cannot use the
approximations used in the previous section, based on small loop momenta
and which conveniently approximated the square root terms by polynomials.
In this case the average loop momenta exceeds even $\mu$. Furthermore, if
one considers the full residues given in section II, one notes that they
contain poles for real positive $k$. The $R_{\1}^{\square}(k)$ residue has
a pole for\[\mu^{\2}-2\,m_{\1}\left(E_{\1}+m_{\1}\right)=0.\] While the
$R_{\0}^{\square}(k)$ residue has one when \[ A = 0 \,\,\,\Rightarrow
\,\,\,2\,E_{\0}m_{\1} =\mu^{\2}. \]  We shall now argue that these pole
contributions cancel. First we observe, the non-obvious fact, that these
singularities occur at the {\em same value} of $k$, i.e. at
\[k_s=\mu\,\sqrt{\left(\frac{\mu}{2\,m_{\1}}\right)^{\2}-1}\,\,\gg \mu . \]
This equation confirms their absence for the case considered in the
previous section, since the value of $k$ at the pole becomes complex for
small $\mu$. On the other hand, there are no pole contributions in the
crossed residues. The cancellation of the box poles can be shown both
analytically and numerically. We will not give here the analytic proof
derived from a Maclaurin series expansion of the box terms about $k_s$.

 The numerical argument is essentially based upon Fig.\,3. In this
figure (drawn for an arbitrary choice of masses compatible with
our limit) we plot the ratio
\[\frac{R_{\1}(k)+R_{\2}(k)}{R_{\0}(k)}
\]
separately for the box and crossed terms  with a change of sign
for the crossed terms for clarity of the figure. The region
plotted in $k$ includes the pole value $k_s$. The curves are
essentially identical. The pole terms in the numerator and
denominator of the box ratio have cancelled resulting in a smooth
curve. Indeed these plots show that there is no observable
difference in the sum of the box and of the crossed residues.
Furthermore, for $k>0.5\,\mu$ the ratio tends rapidly to one or
minus one as the case may be. This means that the the sum of the
three residues cancel not only any pole contributions, but cancel
"tout-court" for $k \geq \mu$. This occurs separately for both the
box and crossed sums and consequently for the total sum. Another
conclusion based upon Fig.\,3 is that for $k\ll\mu$,
\[R_{\1}(k)+R_{\2}(k) \gg  R_{\0}(k)\] separately both for the box
and crossed terms. Numerical trials have lead us to conclude that
integrating $R_{\1}(k)+R_{\2}(k)$ in $k$ up to $\mu/2$ yields an
excellent approximation to the full integration over all three
terms. This is useful, not so much for the numerical calculations
as for the derivation of a closed expression for the box and
crossed diagram contributions. In Table 1 (lower half) we list the
numerical and analytic results based upon the above  heuristic
rule. The agreement is very impressive. The analytic formulas we
used for this table are given below and were derived as follows.
First note that the poles at $k_s$ lie outside our truncated
integrated region ($k<\mu/2$) so that the integrals can be
performed using elementary formulas. We start with the following
simplified expressions for these residues, in which we have
dropped, where possible, the incoming scalar masses compared to
$\mu$,
\begin{eqnarray*}
k^{\2}\,R_{\1}^{\square}(k)  &\approx& \left(E_{\1}-m_{\1}\right)
\,/\,\left[\,4\, W\,E_{\1}\,\mu^{\4}\,\right] \,\,,\\
k^{\2}\,R_{\2}^{\square}(k)  &\approx& -\,\left(E_{\2}+m_{\2}\right)
\,/\,\left[\,4\, W\,E_{\2}\,\mu^{\4}\,\right] \,\,,\\
k^{\2}\,R_{\1}^{\times}(k)  &\approx& \left(E_{\1}+m_{\1}\right)
\,/\,\left[\,4\, W\,E_{\1}\,\mu^{\4}\,\right] \,\,,\\
k^{\2}\,R_{\2}^{\times}(k)  &\approx& -\,\left(E_{\2}+m_{\2}\right)
\,/\,\left[\,4\, W\,E_{\2}\,\mu^{\4}\,\right] \,\,.
\end{eqnarray*}
Consequently,
\begin{eqnarray*}
k^{\2}\,\left[\,R_{\1}^{\square}(k) +R_{\2}^{\square}(k)\,\right] &\approx&
-\,\frac{1}{4\,W\,\mu^{\4}}\,\left(\,\frac{m_{\1}}{E_{\1}} +
\frac{m_{\2}}{E_{\2}} \,\right)\,\,,\\
k^{\2}\,\left[\,R_{\1}^{\times}(k) +R_{\2}^{\times}(k)\,\right] &\approx&
-\,\frac{1}{4\,\Delta\,\mu^{\4}}\,\left(\,\frac{m_{\2}}{E_{\2}} -
\frac{m_{\1}}{E_{\1}} \,\right)\,\,.
\end{eqnarray*}
The integrals up to $\mu/2$ yield,
\begin{eqnarray*}
\int_{\0}^{\mu/2} \mbox{d}k\,\,k^{\2}\,\left[\,R_{\1}^{\square}(k)
+R_{\2}^{\square}(k)\,\right] &\approx&
-\,\frac{1}{4\,W\,\mu^{\3}}\,\left(\,\frac{m_{\1}}{\mu}\,\,
\mbox{arcsinh} \frac{\mu}{2\,m_{\1}}  + \frac{m_{\2}}{\mu}\,\,
\mbox{arcsinh} \frac{\mu}{2\,m_{\2}} \, \right)
\,\,,\\
\int_{\0}^{\mu/2} \mbox{d}k\,\, k^{\2}\,\left[\,R_{\1}^{\times}(k)
+R_{\2}^{\times}(k)\,\right] &\approx&
\,-\,\frac{1}{4\,\Delta\,\mu^{\3}}\,\left(\,\frac{m_{\2}}{\mu}\,\,
\mbox{arcsinh} \frac{\mu}{2\,m_{\2}} - \frac{m_{\1}}{\mu}\,\,
\mbox{arcsinh} \frac{\mu}{2\,m_{\1}} \, \right)\,\,.
\end{eqnarray*}
Finally,
\begin{equation}
\label{th1} \mathcal{M}^{\square} \approx
\frac{\lambda_{\1}^{^{\2}}\lambda_{\2}^{^{\2}}}{2\,\pi^{\2}}\, \left[\,
-\,\frac{1}{4\,W\,\mu^{\3}}\,\left(\,\frac{m_{\1}}{\mu}\,\, \mbox{arcsinh}
\frac{\mu}{2\,m_{\1}}  + \frac{m_{\2}}{\mu}\,\, \mbox{arcsinh}
\frac{\mu}{2\,m_{\2}} \, \right)
 \right]
\end{equation}
and
\begin{equation}
\label{th2} \mathcal{M}^{\times} \approx
\frac{\lambda_{\1}^{^{\2}}\lambda_{\2}^{^{\2}}}{2\,\pi^{\2}}\, \left[\,
-\,\frac{1}{4\,\Delta\,\mu^{\3}}\,\left(\,\frac{m_{\2}}{\mu}\,\,
\mbox{arcsinh} \frac{\mu}{2\,m_{\2}} - \frac{m_{\1}}{\mu}\,\,
\mbox{arcsinh} \frac{\mu}{2\,m_{\1}} \, \right) \right]\,\,.
\end{equation}
In the lower half of Table 1 (high mass exchange) by "analytic
box" and "analytic crossed", we intend the expressions in the
brackets of Eqs.(\ref{th1}) and (\ref{th2}). The numerical
calculations have been made for the sum of all the three residues
and without an explicit cut-off in $k$.

For a more compact expression, we now add these "analytic" results
after approximating the $\mbox{arcsinh}(x/2)$ by $\ln x$, since
all our $x$ are very large. After some algebra, we obtain the
following formula for the fourth order contributions to the
invariant amplitude,
\begin{eqnarray}
\mathcal{M}^{\square}+\mathcal{M}^{\times} & \approx &
\frac{\lambda_{\1}^{^{\2}}\lambda_{\2}^{^{\2}}}{2\,\pi^{\2}}\, \left[\,-
\,\frac{1}{4\,\mu^{\4}}\,\left( \,\ln\frac{\mu^{\2}}{m_{\1}m_{\2}} +
\frac{1+\rho^{\2}}{1-\rho^{\2}} \ln \rho\,\right) \,
 \right] \nonumber\\
  & = &
  -\,\frac{\lambda_{\1}^{^{\2}}\lambda_{\2}^{^{\2}}}{8\,\pi^{^{\2}}}\,
  \frac{1}{\mu^{\4}}\,
\left( \,\ln\frac{\mu^{\2}}{m_{\1}m_{\2}} + \frac{1+\rho^{\2}}{1-\rho^{\2}}
\ln \rho\,\right)\,\,,
\end{eqnarray}
with $\rho=m_{\1}/m_{\2}$. As an aside we note that this result is
symmetric in the incoming masses $m_{\1}$ and $m_{\2}$. This
natural result is not obvious in the expressions for the fourth
order diagrams. The condition for the existence for a bound state
in the high mass exchange case thus becomes
\begin{equation}
\frac{g^{\2}_{\eff}}{2\,\pi^{^{\2}}}  \geq  \frac{\mu^{\2}}{m_{\1}m_{\2}}
\,\mbox{\large /}\,\left( \,\ln\frac{\mu^{\2}}{m_{\1}m_{\2}} +
\frac{1+\rho^{\2}}{1-\rho^{\2}} \ln \rho\,\right)\,\,.
\end{equation}
This inequality is even simpler when $m_{\1}\ll m_{\2}$. In this
case one obtains
\begin{eqnarray*}
\left(\,\ln\frac{\mu^{\2}}{m_{\1}m_{\2}} + \frac{1+\rho^{\2}}{1-\rho^{\2}}
\ln \rho\,\right)\to\,2\ln\frac{\mu}{m_{\2}}\,\,,
\end{eqnarray*}
and hence for a bound state to exist (when $ m_{\1}\ll m_{\2} \ll
\mu $) one must have
\begin{equation}
g_{\eff}^{\2} \geq
\pi^{\2}\,\displaystyle{\frac{\mu}{m_{\1}}\,\frac{\mu/m_{\2}}{\ln
[\mu/m_{\2}]}}\,\,.
\end{equation}
Since $\mu/m_{\2}\gg\ln[\mu/m_{\2}]$, this is an even stronger condition on
the coupling strength than the low mass exchange condition, extrapolated to
high mass exchange,
\begin{equation}
g_{\eff}^{\2}  \geq 4\,\pi\,\displaystyle{\frac{\mu}{m_{\1}}}\,\,.
\end{equation}
 We conclude that, in our toy
model, bound states for high mass exchanges do not exist unless
the effective coupling constant becomes even stronger than that
required by the Hulthen condition.

\section*{V CONCLUSIONS}

We have presented in this paper a calculation of the forward
scattering contributions of the fourth order box and lader
diagrams for a particular scalar field model. From these results
the condition upon the effective coupling strength for the
existence of a bound state has been obtained. The requirement
imposed was that the sum of these  fourth order terms equal or
exceed the tree diagram contribution. For small exchanged mass
($\mu \ll m_{\1,\2}$) we have reobtained the result of Franz
Gross\cite{GROSS,G69}. In the opposite limit of high exchanged
mass ($\mu \gg m_{\1,\2}$) we have derived an inequality for a
bound state, albeit as an approximate result. It agrees very well
 with our numerical integral results for appropriate (but
otherwise casually chosen) sets of selected mass values.

A first observation to be made is that the two inequalities, for
low and high $\mu/m$, are not the same. One should therefore not
extrapolate either outside of their respective domains. In this
particular model, the conclusion is that, as the exchanged mass
increases, the effective coupling constant must grow even faster
than $\sqrt{\mu/m}$ for a bound state to exist. This is a toy
model so we have no explicit (physical) limitations, but of course
large coupling constants negate the very perturbation series upon
which the method is based. However, this is not what we expect to
happen for interacting spinors. It is perhaps useful to recall
here, more explicitly, some of the arguments upon which our
expectations for spinors are based. Amongst the lowest order
relativistic corrections to the Schroedinger equation is that
which gives rise to the renowned Darwin term \cite{SAK},
\[\frac{e}{8m^{\2}}\,\nabla^{^{\2}}\,A_{\0}(r)\,\,.\]
This term simply adds onto the potential term $eA_{\0}(r)$. When the
electrostatic potential is a Coulomb potential produced by an opposite
charged point (massive) source, we obtain
\[ \frac{e^{\2}}{8m^{\2}}\,\delta^3(\boldsymbol{r})\,\,.\]
This contributes only to the $s$\,-wave but it is essential for
the transformation of the
 relativistic correction of the Hydrogen energy spectrum into
 one which depends only upon $j$ (the total angular
 momentum) in addition to $n$ (the principle quantum number).
 A result which is automatic in the Dirac equation. We
note that since the Darwin term is essential to the s-wave spinor
bound
 states, these are technically "relativistic"
 under our definition (see the Introduction).
 However, when this same term is calculated for a Yukawa potential, we
observe that
\[ \nabla^{^{\2}}\,\frac{\exp[-\,\mu\,r]}{r} = \mu^{\2}\, \frac{\exp[-\,\mu\,r]}{r}
-4 \,\pi\,\delta^3(\boldsymbol{r})\,\,. \]
The first of these terms
augments the Yukawa potential and
 amplifies the effective coupling constant,
\[ g_{\eff}^{\2} \to g_{\eff}^{\2}\,\left(1+\frac{\mu^{\2}}{8m^{\2}}\right)
 \,\,.\]
 For $\mu\gg m$,
\[ g_{\eff}^{\2} \to g_{\eff}^{\2}\,\frac{\mu^{\2}}{8m^{\2}}
 \,\,,\]
 and this is just just what is needed to "invert" the inequality condition for a bound
state from
\[
g_{\eff}^2 > \frac{\mu}{m}\,\,\, \to \,\,\,g_{\eff}^{\2} >
\frac{8m}{\mu}\,\,.
\]
The latter inequality is a weak constraint, easily satisfied, since $m \ll
\mu$. Of course, this argument is flawed by the fact that limiting oneself
to the lowest order relativistic corrections assumes that they must be
small, or at least that the higher order corrections can for some reason
(such as cancellations) be totally ignored. Nevertheless, this result does
suggest that relativistic effects could be very important for the bound
state inequality. In the specific model treated in this paper Gross has
shown that the relativistic corrections for small $\mu/m$ come from the
poles in the double pole contributions\cite{GROSS}. Now the box and crossed
contributions for these double poles cancel in this model. It is therefore
a very different situation from the case of interacting spinors.
Furthermore, the Klein-Gordon equation does not have a Darwin type term, so
Yukawa coupling amplification has not been shown to occur for interacting
scalars.  On the contrary, the results of this paper demonstrate
specifically that it does not occur. It is our intention to consider a more
interesting model with incoming spinors exchanging bosons in a future
study.

A possible alternative approach in determining the inequality
condition for a bound state is to first derive a corresponding two
body differential equation (Bethe-Salpeter in this casa of scalar
interactions) from which not only the existence of a bound state
may be derived but indeed the full bound state spectrum. However,
our procedure is the only one available for cases in which the
two-body equation is unknown\cite{GROSS}.

\newpage

\begin{table}
\begin{center}
\begin{tabular}{|l|l|l||l|l||l|l|} \hline
 & & & & & & \\
$m_{\1}$ & $m_{\2}$ & $\mu$ & $\begin{array}{ll} \mbox{Analytic} \\
\mbox{Box}  \\
\\ \left[\, \lambda_{\1}^{\2}\lambda_{\2}^{\2}\,/\,2\,\pi \,\right] \end{array}$ &
$\begin{array}{ll} \mbox{Numerical} \\ \mbox{Box}  \\  \\\left[\,
\lambda_{\1}^{\2}\lambda_{\2}^{\2}\,/\,2\,\pi \,\right]
\end{array}$ &
$\begin{array}{ll} \mbox{Analytic} \\ \mbox{Crossed}  \\  \\ \left[\,
\lambda_{\1}^{\2}\lambda_{\2}^{\2}\,/\,2\,\pi^{\2} \,\right]\end{array}$  &
\ $\begin{array}{ll} \mbox{Numerical} \\  \mbox{Crossed}  \\ \\
\left[\, \lambda_{\1}^{\2}\lambda_{\2}^{\2}\,/\,2\,\pi^{\2} \,\right] \end{array}$  \\
& & & & & & \\
\hline \hline & & & & & & \\
$1$ & $2$ & $10^{-\3}$ &$-\,1.3087\times 10^{\8}$ & $-\,1.3087\times
10^{\8}$&
$-\,3.1250 \times 10^{\4}$ & $-\,3.1232 \times 10^{\4}$ \\& & & & & & \\  \hline & & & & & & \\
$1$ & $4$ & $10^{-\3}$ & $-\,7.8524 \times 10^{\7}$ &$-\,7.8524 \times
10^{\7}$ &
$-\,1.5625 \times 10^{\4}$ & $-\,1.5617 \times 10^{\4}$  \\ & & &  & & & \\ \hline & & & & & & \\
$2$ & $4$ & $10^{-\3}$ & $-\,6.5442\times 10^{\7}$ & $-\,6.5442\times
10^{\7}$& $ -\,7.8125 \times 10^{\3}$ & $-\,7.8104 \times 10^{\3}$ \\ & & & & & & \\ \hline & & & & & & \\
$1$ & $2$ & $10^{-\2}$ & $-\,1.3059 \times 10^{\5}$ & $-\,1.3059 \times
10^{\5}$ & $-\,3.1250 \times 10^{\2}$&
$-\,3.1067 \times 10^{\2}$\\
 & & & & & & \\
\hline
\hline & & & & & & \\
$10^{-\2}$ & $2 \times 10^{-\2}$ & $1$ &$-\,1.0358$ & $-\,1.0363$&
$-\,0.8049$&
$-\,0.8059$ \\& & & & & & \\ \hline & & & & & & \\
$10^{-\2}$ & $4\times 10^{-\2}$ & $10$ &$-\,1.4497 \times 10^{-\4}$ &
$-\,1.4497 \times 10^{-\4}$&
$-\,1.2648 \times 10^{-\4}$ &$-\,1.2649 \times 10^{-\4}$ \\ & & & & & & \\ \hline & & & & & & \\
$2\times 10^{-\2}$ & $4\times 10^{-\2}$ & $10$ &$-\,1.4381 \times 10^{-\4}$
&$-\,1.4382 \times 10^{-\4}$ &
$-\,1.2071 \times 10^{-\4}$ & $-\,1.2071 \times 10^{-\4}$\\ & & & & & & \\
\hline & & & & & & \\
$10^{-\2}$ & $2\times 10^{-\2}$ & $10$ &$-\,1.6114 \times 10^{-\4}$
&$-\,1.6114 \times 10^{-\4}$ &
$-\,1.3804 \times 10^{-\4}$ &$-\,1.3804 \times 10^{-\4}$ \\ & & & & & & \\
\hline
\end{tabular}
\end{center}
\caption{We list the numerical and analytic results for the exchange of
small (upper half) and high (lower half) mass scalars. The analytic
formulas  for the small $\mu$ limit refer for the box diagram to the
expression in the brackets of  Eq.(\ref{ts1}) and for the crossed one to
that of Eq.(\ref{ts2}). The analytic formulas  for the high $\mu$ limit
refer to Eqs. (\ref{th1}) and (\ref{th2}).}
\end{table}

\newpage

\begin{figure}[hbp]
\hspace*{-2.5cm}
\includegraphics[width=19cm, height=24cm, angle=0]{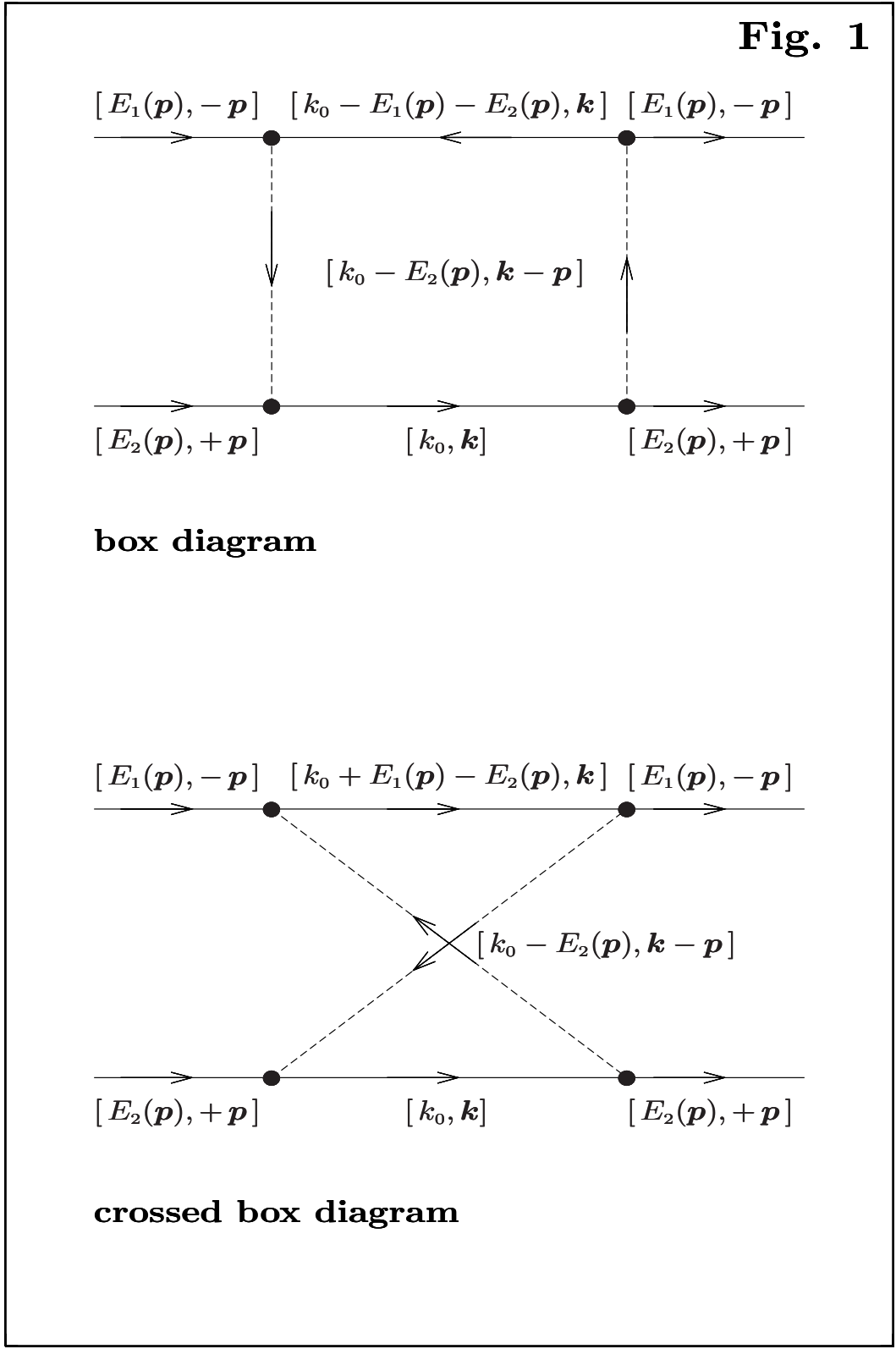}
\vspace*{-2cm}
 \caption{The fourth order box and crossed box diagrams in a scalar field
 model evaluated in the center of mass frame for scattering in the forward
direction.}
\end{figure}

\newpage

\begin{figure}[hbp]
\hspace*{-2.5cm}
\includegraphics[width=19cm, height=24cm, angle=0]{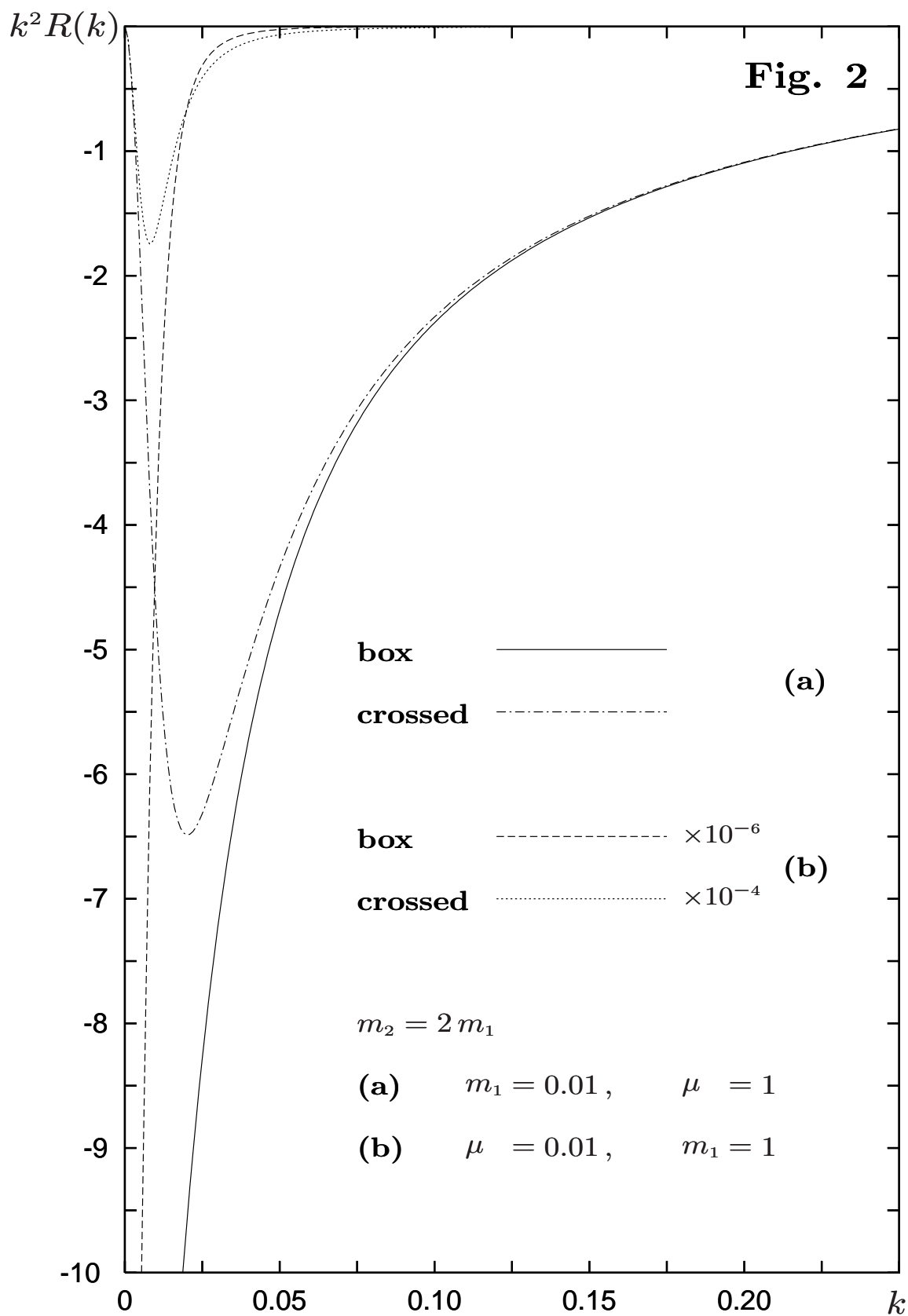}
\vspace*{-2cm}
 \caption{The $k$-dependence of the box and crossed residues
 are plotted for particular values of $m_{\1,\2}$ and $\mu$. The case (a) and (b)
correspond respectively to the exchange of high and small mass
scalars.}
\end{figure}

\newpage

\begin{figure}[hbp]
\hspace*{-2.5cm}
\includegraphics[width=19cm, height=24cm, angle=0]{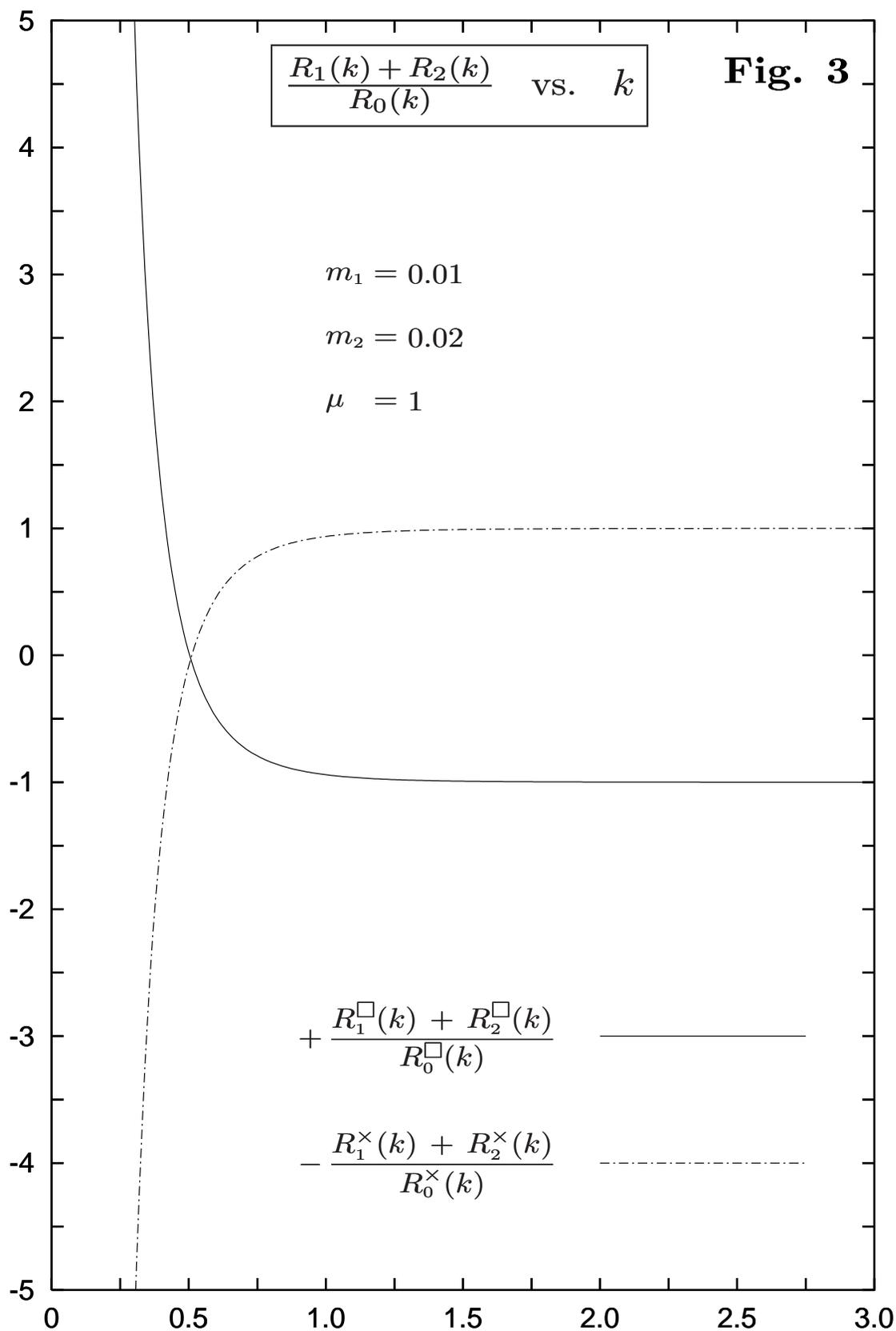}
\vspace*{-2cm}
 \caption{The plots refer to the ratio of $R_{\1}(k)$ plus $R_{\2}(k)$ over
 $R_{\0}(k)$ for the box and crossed terms. For clarity the crossed terms are
 plotted with a change of sign. The curves are essentially identical and for
 $k\geq \mu$ the sum of the three residues cancel.}
\end{figure}

\end{document}